# QUASI-STATIC AND DYNAMIC BEHAVIOR OF ADDITIVELY MANUFACTURED METALLIC LATTICE CYLINDERS[*]

Hossein Sadeghi[1†], Dhruv Bhate[2], Joseph Abraham[1], Joseph Magallanes[1]

[1]Karagozian & Case, 700 N. Brand Blvd Ste 700, Glendale, CA 91203
[2]The Polytechnic School, Arizona State University, 7001 E Williams Field Rd, Mesa, AZ 85212

## ABSTRACT

Lattice structures have tailorable mechanical properties which allows them to exhibit superior mechanical properties (per unit weight) beyond what is achievable through natural materials. In this paper, quasi-static and dynamic behavior of additively manufactured stainless steel lattice cylinders is studied. Cylindrical samples with internal lattice structure are fabricated by a laser powder bed fusion system. Equivalent hollow cylindrical samples with the same length, outer diameter, and mass (larger wall thickness) are also fabricated. Split Hopkinson bar is used to study the behavior of the specimens under high strain rate loading. It is observed that lattice cylinders reduce the transmitted wave amplitude up to about 21% compared to their equivalent hollow cylinders. However, the lower transmitted wave energy in lattice cylinders comes at the expense of a greater reduction in their stiffness, when compared to their equivalent hollow cylinder. In addition, it is observed that increasing the loading rate by five orders of magnitude leads to up to about 36% increase in the peak force that the lattice cylinder can carry, which is attributed to strain rate hardening effect in the bulk stainless steel material. Finite element simulations of the specimens under dynamic loads are performed to study the effect of strain rate hardening, thermal softening, and the failure mode on dynamic behavior of the specimens. Numerical results are compared with experimental data and good qualitative agreement is observed.

## 1 Introduction

Due to recent advances in additive manufacturing (AM), parts with complex geometries which were not possible (or very difficult) to fabricate through traditional manufacturing techniques, can now be fabricated [1]. This allows engineers to use topology optimization to design complex parts with enhanced mechanical performance and reduced weight, and to make them through AM. Specifically, AM has enabled fabrication of architected lattice structures to make parts with tuned mechanical properties [2, 3, 4]. Of particular interest is the capability of lattices for tailoring shock wave propagation and impact energy absorption to minimize the transmitted wave energy to a structure [5, 6, 7]. To maximize the shock and impact energy absorption, the architecture of lattices should be optimally designed to employ various energy dissipation mechanisms, i.e. plastic deformation, buckling, transforming the axial load to lateral deformation, etc. However, the mechanical performance of lattice structures made by AM under dynamic loads is not well-understood and further studies are needed to characterize such structures at various loading rates.

---

Wadley et al. [5] studied the compressive response of multilayered pyramidal lattices during underwater shock loading. They observed that the lattice structure crushes in a progressive manner by the sequential buckling of the struts, which leads to extended duration of the recorded back-face pressure-time waveform and significant reduction in the transmitted pressure measured at the back-face. Mckown et al. [6] studied the failure mechanisms of additively manufactured lattice structures under quasi-static and blast loadings. Buckling failure was observed in some of the lattice structures, whereas, a progressive mode of collapse was evident in other samples. They observed a rate sensitivity of the yield strength of the lattice structures as well as an increase in the blast resistance of the lattice structures with increasing the yield strength. Smith et al. [8] performed quasi-static and blast tests on additively manufactured lattice structures made of stainless steel 316L. They observed that there is significant increase in the lattice yield strength at high strain rates which improves the energy absorption of the lattices and makes them more capable of withstanding blast loads. They also observed both the progressive damage and buckling-dominated modes of collapse in lattices with different architecture. Hawreliac et al. [7] studied dynamic behavior of additively manufactured lattice structures under impact loading. Using in situ x-ray phase contrast imaging, they observed elastic wave propagation in the lattice structures for both bend and stretch dominated lattices. They concluded that the elastic properties of periodic lattice structures can be predicted accurately by the Bloch wave theory. Ozdemir et al. [9] studied energy absorption of lattice structures under quasi-static and dynamic loading. Cubic, diamond, and reentrant cube additively manufactured lattice structures were tested under quasi-static and dynamic loads and it was observed that lattice structures are able to spread impact loading in time and to reduce the peak impact stress. Tancogne-Dejean et al. [10] studied energy absorption of additively manufactured stainless steel 316L octet lattice structures under quasi-static and dynamic loading. They observed that the specific energy absorption of lattice structures is a monotonically increasing function of their relative density. In addition, they reported that the overall strength of the lattice structures increases by increasing the strain rate.

Lattices are most beneficial when used in sandwich structures, i.e. two thin and stiff skins (e.g. FRP or metal) with a light and energy-absorptive core (e.g. foam or lattice), to provide high in-plane (bending) stiffness as well as high through-the-thickness energy absorption. Thin- and thick-wall cylinders are commonly used in various industries, e.g. aerospace, defense, oil and gas, etc., and there is a need to study the use of sandwich structures with cylindrical skin and internal lattice core. Especially, due to the weight limitations in the aerospace applications, there is a demand for cylindrical structures with maximum energy absorption and reduced weight. Furthermore, an inherent limitation of the natural materials is that, high energy absorption usually comes at the expense of lowering the material stiffness and strength. Although, there have been several studies on the benefits provided by lattices, there is no study that compares the performance of a lattice structure with an equivalent structure with no lattice, as far as the authors know. In this paper, quasi-static and dynamic behavior of additively manufactured stainless steel cylindrical samples with internal lattice structure is studied. Equivalent hollow cylindrical samples with the same length, outer diameter, and mass (larger wall thickness) are also fabricated. Samples are tested under high rate deformation using a split Hopkinson bar, as well as at quasi-static regime. Finite element simulation of the specimens under dynamic loads are performed and the results are compared with experimental measurements.

## 2 Experimental Procedure

### 2.1 Specimen Fabrication

Figure 1(a, b) show schematic drawings of a cylinder with internal lattice structure (lattice cylinder) and its equivalent hollow cylinder (HC) with the same length, outer diameter, and mass (larger wall thickness), respectively. The unit cell of the lattice and a section cut of the lattice cylinder (LC) are shown in Figure 1(c, d), respectively. Three different lattice cylinders (LC) with different cylinder wall thickness and strut diameter and their equivalent HC specimens are fabricated, where their dimensions are given in Table 1. The strut length for all the LC specimens is 3 mm. The metal 3D printer used in this study is a Concept Laser MLab Cusing R machine which is a laser powder bed fusion system. The powder material used for fabrication of the specimens is stainless steel 316L and its composition is given in Table 2. A layer thickness of 25 µm and spot diameter of 50 µm is used in the fabrication process. Solid cylinders (SC) are also fabricated by AM to measure the bulk material properties of the additively manufactured stainless steel 316L. The diameter and length of the SC specimens for SHB and quasi-static tests are given by $d_{\text{SHB}} = 2.97$ mm, $L_{\text{SHB}} = 2.98$ mm; and $d_{\text{qs}} = 6.00$ mm, $L_{\text{qs}} = 8.76$ mm, respectively.

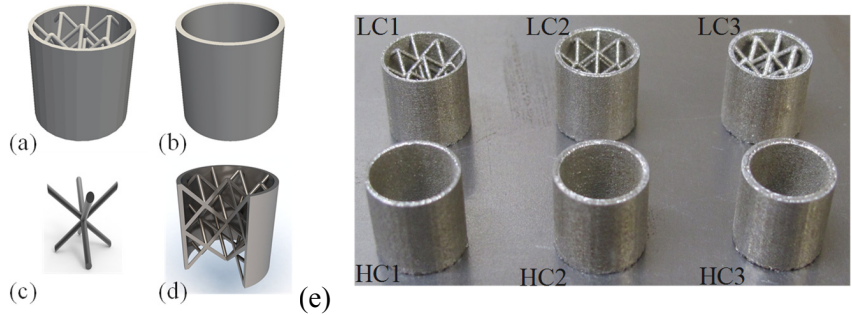

Figure 1: Schematic drawings of (a) lattice cylinder (LC), (b) hollow cylinder (HC), (c) lattice unit cell, (d) section cut of the LC, and (e) photograph of the specimens.

Table 1: Specimens dimensions in millimeter.

|  | Abr. | Length | Diameter | Wall Thick. | Strut Dia. |
|---|---|---|---|---|---|
| Lattice Cylinder 1 | LC1 | 8.55 | 8.55 | 0.36 | 0.36 |
| Lattice Cylinder 2 | LC2 | 8.55 | 8.55 | 0.54 | 0.36 |
| Lattice Cylinder 3 | LC3 | 8.55 | 8.55 | 0.54 | 0.54 |
| Hollow Cylinder 1 | HC1 | 8.55 | 8.55 | 0.45 | - |
| Hollow Cylinder 2 | HC2 | 8.55 | 8.55 | 0.63 | - |
| Hollow Cylinder 3 | HC3 | 8.55 | 8.55 | 0.74 | - |

Table 2: Chemical composition of stainless steel 316L powder used in the specimen fabrication.

| Component | Fe | Cr | Ni | Mo | Mn | Si | P | C | S |
|---|---|---|---|---|---|---|---|---|---|
| Percentage | Balance | 16.5-18.5 | 10-13 | 2-2.5 | 0-2 | 0-1 | 0-0.045 | 0-0.03 | 0-0.03 |

### 2.2 Quasi-static Test

Figure 2 shows a photograph of the MTS 810 (20 kips) quasi-static test set-up used in this study. All the quasi-static tests are performed at the velocity of 0.001 in/sec and each test is performed twice to ensure repeatability.

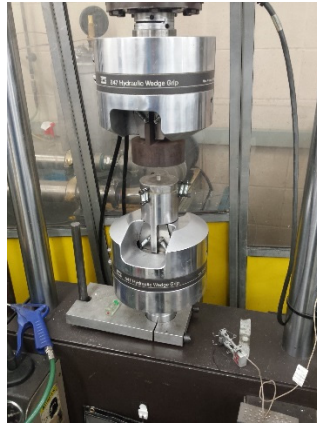
Figure 2: Photograph of the quasi-static test setup (MTS 810).

## 2.3  *Split Hopkinson Bar Test*

Split Hopkinson bar (SHB) is widely used for characterization of materials under plastic deformation at high strain rates. Figure 3(a, b) show a schematic drawing and a photograph of the SHB equipment used in this study. A gas gun sends a projectile with a desired velocity towards the end A of the incident bar, which in turn creates a compressive stress wave propagating toward the sample. The sample is sandwiched between end B of the incident bar and end C of the transmission bar. When the stress wave reaches the sample, a portion of the pulse is transmitted to the transmission bar, and a portion of it is reflected back into the incident bar. Strain gauges, S1 and S2, measure the strain in the middle of the incident and transmission bars as a function of time. The strain rate, $\dot{\varepsilon}(t)$, strain, $\varepsilon(t)$, and stress, $\sigma(t)$, in the specimen can be calculate by [11]

$$\dot{\varepsilon}(t) = -2\frac{c_0}{L}\varepsilon_R(t)$$
$$\varepsilon(t) = \int_0^t \dot{\varepsilon}(\tau)\,d\tau \qquad (1)$$
$$\sigma(t) = \frac{A_0}{A}E\varepsilon_T(t)$$

where $\varepsilon_R$ is the reflected strain, $\varepsilon_T$ is the transmitted strain, $c_0$ is the bar wave velocity, $L$ is the specimen length, $A$ is the specimen cross-sectional area, $A_0$ is the bar cross-sectional area, and $E$ is the bar Young's modulus. Figure 3(c) shows a typical signal recorded by the strain gauges in the SHB equipment.

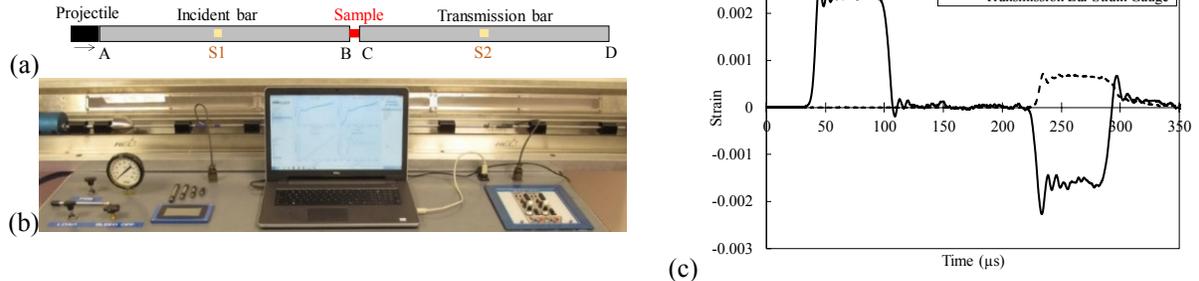

Figure 3: (a) Schematic drawing and (b) a photograph of the SHB equipment, and (c) a typical signal measured by the SHB equipment.

## 3 Finite Element Analysis

Finite element (FE) model of the LC and HC specimens is created in the commercial finite element software LS-DYNA. Due to complexity of the LC geometry, open source auto-mesh generation library CGAL [12] is used to mesh the LC geometry using tetrahedron elements (see Figure 4(a)). One-point nodal pressure tetrahedron elements with alleviated volumetric locking (ELFORM=13) are used for creating the LC models. Constant stress solid hexahedra elements (ELFORM=1) are used to create the HC models. Element size in each model is chosen so there is a minimum of three elements through the cylinder thickness and lattice strut diameter. Due to small size of the elements in the specimen it was impractical (computational expensive) to model the SHB bars explicitly, therefore, only two plates with thickness of 1 mm are modeled on the top and bottom of the specimens to model the interaction of the specimen with the bars, as shown in Figure 4(b). The contact between the bars and plates is modeled with automatic surface to surface contact with friction. The bottom plate is kept fixed and the velocity of the top plate is prescribed to have a profile similar to the incident wave particle velocity measured by the strain gauge in the SHB test, as show in Figure 4(c). Johnson-Cook (JC) material model [13], which has the following form, is used in the simulations

$$\sigma = (A + B\varepsilon^n)(1 + Cln(\dot{\varepsilon}^*))(1 - T^{*m}) \qquad (2)$$

where $\dot{\varepsilon}^*$ is the normalized strain rate, $T^*$ is normalized temperature, and $A, B, n, C$ and $m$ are material parameters. In this phenological material model, the terms in the parentheses describe the strain hardening, strain rate hardening, and thermal softening behavior of the material, respectively.

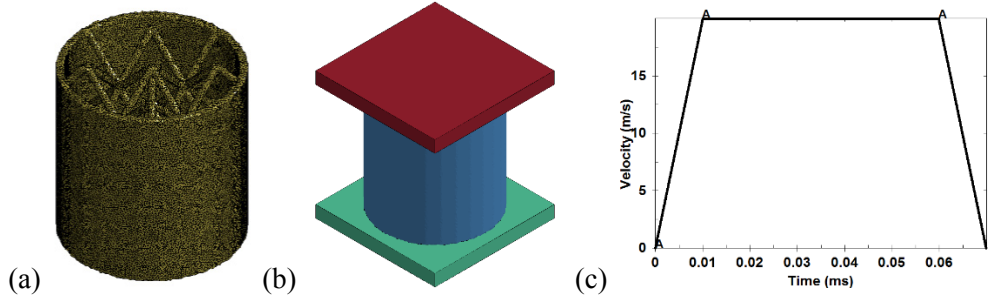

Figure 4: (a) FE model of LC1 specimen, (b) schematic drawing of the specimen between the top and bottom plates, and (c) prescribed top plate velocity.

## 4 Results

### 4.1 AM-SS316L Bulk Material Properties

Figure 5 shows the quasi-static (QS) and dynamic (SHB) true stress-strain curves of the bulk additively manufactured stainless steel 316L (AM-SS316L) at strain rates of 0.01 1/s and 4300 1/s, respectively. A picture of the quasi-static and SHB SC specimens after the tests are also shown in this figure. Table 3 shows the fitted JC material parameters for the AM-SS316L together with the JC parameters for traditionally manufactured SS316L for comparison. These material parameters are used to create a JC material model for used in the FE simulations. A comparison of the stress-strain curve calculated by JC model with the test data is also given in Figure 5. It should be noted that the thermal softening coefficient, *m,* is not measured in this study.

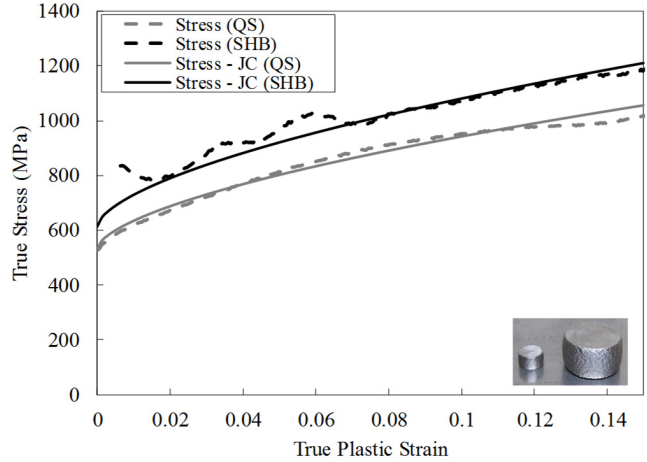

Figure 5: Quasi-static and dynamic true stress-strain curves for additively manufactured stainless steel 316L at strain rates of 0.01 1/s and 4300 1/s.

Table 3: JC parameters for AM-SS316L and traditionally manufactured stainless steel.

|  | A (MPa) | B (MPa) | n | C | m | $\dot{\varepsilon}_0$ (1/s) |
|---|---|---|---|---|---|---|
| AM-SS316L | 534 | 1641 | 0.604 | 0.011 | - | 0.01 |
| Changeux et al. [14] | 280 | 1750 | 0.8 | 0.1 | 0.85 | 200 |
| Tounsi et al. [15] | 514 | 514 | 0.508 | 0.042 | 0.533 | 0.001 |

## 4.2 Quasi-static Test

Figure 5(a-c) show the quasi-static force-displacement curves for different specimens as well as a photograph of the specimens after the test. All the curves show a relatively linear behavior up to a point where the material yields and plastic deformation initiates. Due to strain hardening effect, the forces keep increasing after the yield point until it reaches a peak value, where the buckling initiates. After the buckling occurs, the material softens (post-buckling softening) until cylinder or struts self-contact initiates, where the forces start to increase again. It can be seen that the peak force in all LC specimens is lower compared to their equivalent HC specimens. In addition, the stiffness (slope of the linear part of the curve) of the LC specimens is lower than their equivalent HC specimens. It can also be observed that the buckling in LC specimens initiates at a slightly lower displacement compared to their equivalent HC specimens. Also, as expected, the peak force where the buckling initiates occurs at higher displacements when comparing specimen set 1 (LC1 and HC1) with specimen set 2 (LC2 and HC2), and specimen set 2 with specimen set 3 (LC3 and HC3). It is interesting to see that the deformation in all specimens is localized at one end. This behavior will be studied and explained later by FE simulations.

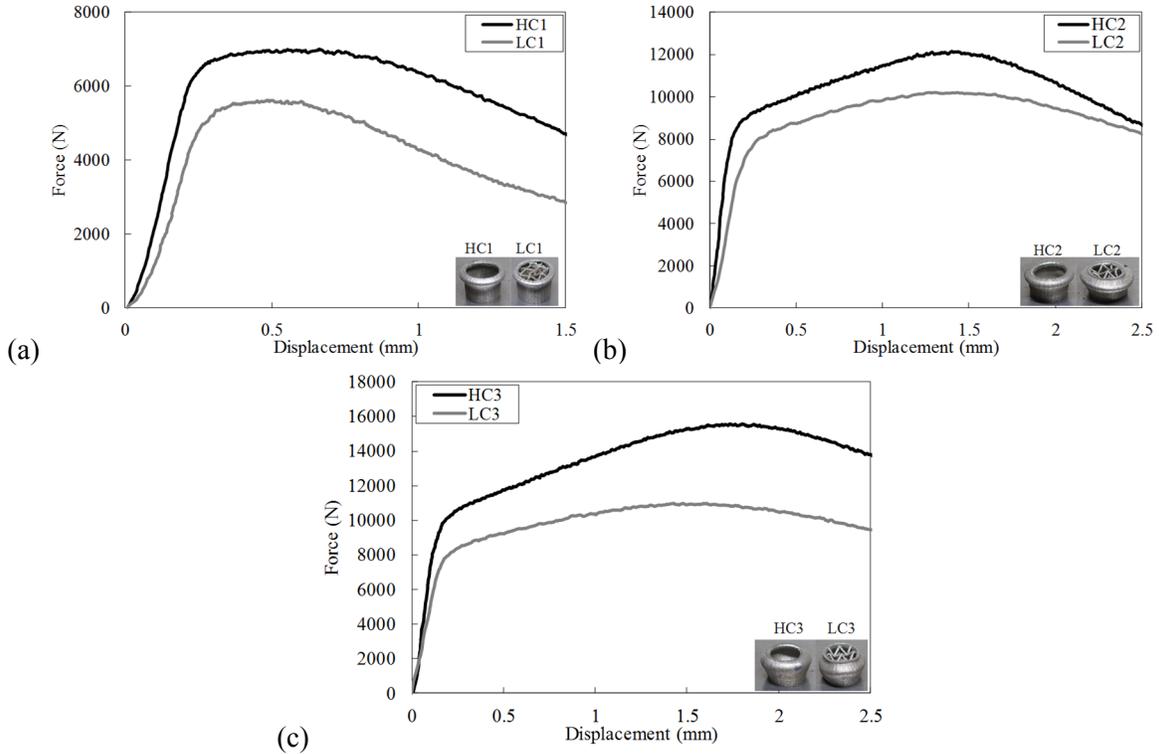

Figure 6: Quasi-static force-displacement curves for (a) HC1 and LC1, (b) HC2 and LC2, and (c) HC3 and LC3 specimens.

*4.3    Split Hopkinson Bar Test*

Figure 1 shows dynamic transmitted force-displacement curves for different specimens measured by SHB equipment. It can be seen that the dynamic force-displacement curves show a similar qualitative behavior as their corresponding quasi-static curves. In particular, similar to what was observed in the quasi-static tests, HC1 and LC1 specimens exhibit softening behavior associated with specimen buckling at approximately 0.5 mm displacement. In addition, the measured dynamic forces for HC2, LC2, HC3 and LC3 specimens keep increasing (within the displacement range measured by SHB) which is in agreement with quasi-static test results. Furthermore, it is observed that lattice cylinders reduce the transmitted wave amplitude up to 21% compared to their equivalent hollow cylinders. Displacement rate is also shown in Figure 1 which exhibit a near constant value after the plastic deformation occurs. Similar to quasi-static test results, the deformation in specimen set # 1 is localized at one end; while, for specimen set # 2 and 3 the deformation is (relatively) uniform. The localized deformation in LC1 and HC1 specimens is mainly in the form of cylinder and strut buckling, while the incident stress wave is not high enough to create buckling in specimen set # 2 and 3. In addition, although, LC1 exhibits post-buckling softening behavior at slightly lower displacement than HC1, the drop in the transmitted force for LC1 specimen is gradual, while for HC1 there is a sudden drop in the transmitted force at about 0.7 mm displacement. This shows that although the lattice structure promotes buckling in LC1 specimen, the struts allow for the buckling to occur in a more controlled manner. Table 4 shows the specimen length and average diameter after the SHB tests. It can be seen that the average diameter of the LC specimens is larger than their equivalent HC specimens. This indicates that the lattice structure transforms energy of the axial incident wave into lateral deformation of the LC specimens, leading to a greater reduction in the transmitted force compared to their equivalent HC specimens. This behavior leads to the lattice structure promoting buckling in the LC specimens, which is in agreement with what is observed in the force-displacement curves.

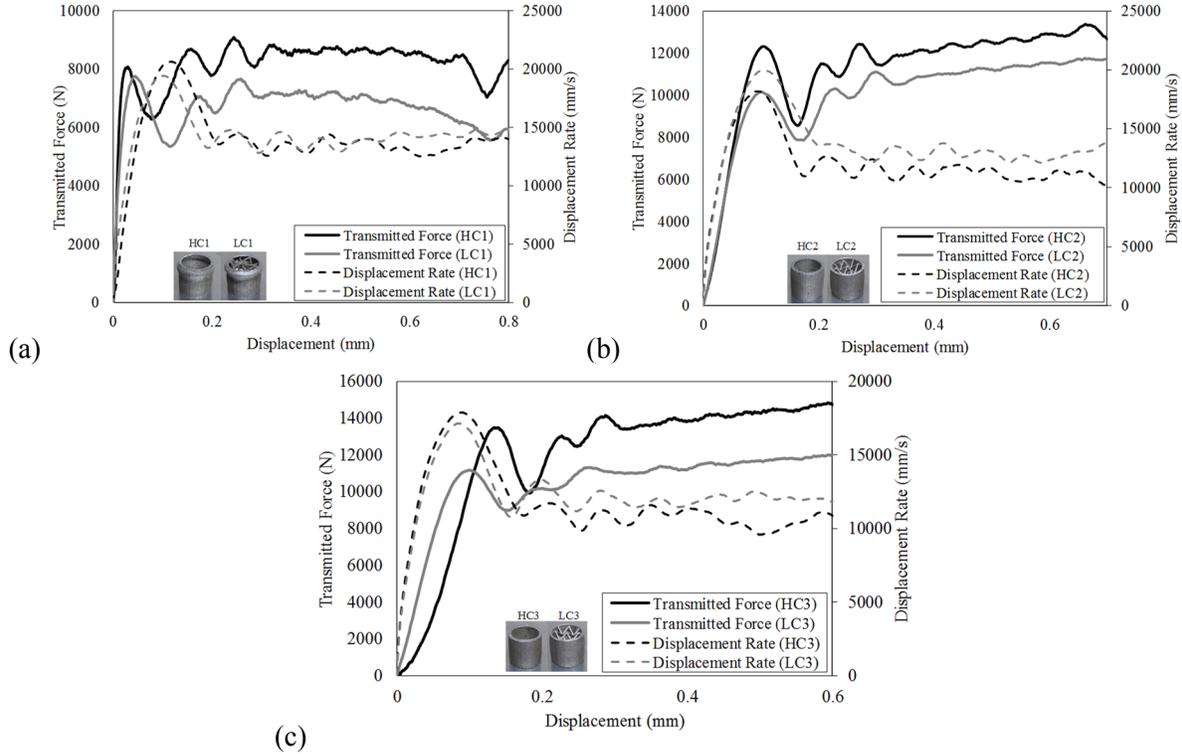

Figure 7: Dynamic force-displacement curves for (a) HC1 and LC1, (b) HC2 and LC2, and (c) HC3 and LC3 specimens.

Table 4: Specimens dimensions in millimeters after the SHB test.

|  | HC1 | HC2 | HC3 | LC1 | LC2 | LC3 |
| --- | --- | --- | --- | --- | --- | --- |
| Length | 7.82 | 8.04 | 8.18 | 7.41 | 7.95 | 8.03 |
| Avg. diameter | 9.15 | 8.84 | 8.75 | 9.34 | 8.99 | 8.97 |

Quasi-static test is a slow process and force equilibrium in the specimens is guaranteed during the deformation, however, dynamic SHB test is a rapid process and the force equilibrium should be studied. Force equilibrium shows that specimen deformation is nearly uniform during the test, while, imbalance of forces could cause the deformation in the sample to be nonuniform (local) and progressive damage could occur. Figure 8 shows the forces at the front and back faces of the specimen during the SHB test for all specimens. It can be seen that although specimen set # 2 and 3 are more or less in equilibrium, specimen set # 1 is not. Specifically, the forces at the front-face of the HC1 and LC1 specimen is higher than the forces in the back-face, which is an indication of localized deformation. The progressive damage behavior in lattices has been previously observed in other studies [9, 5] which are also associated with imbalance of forces in the lattice.

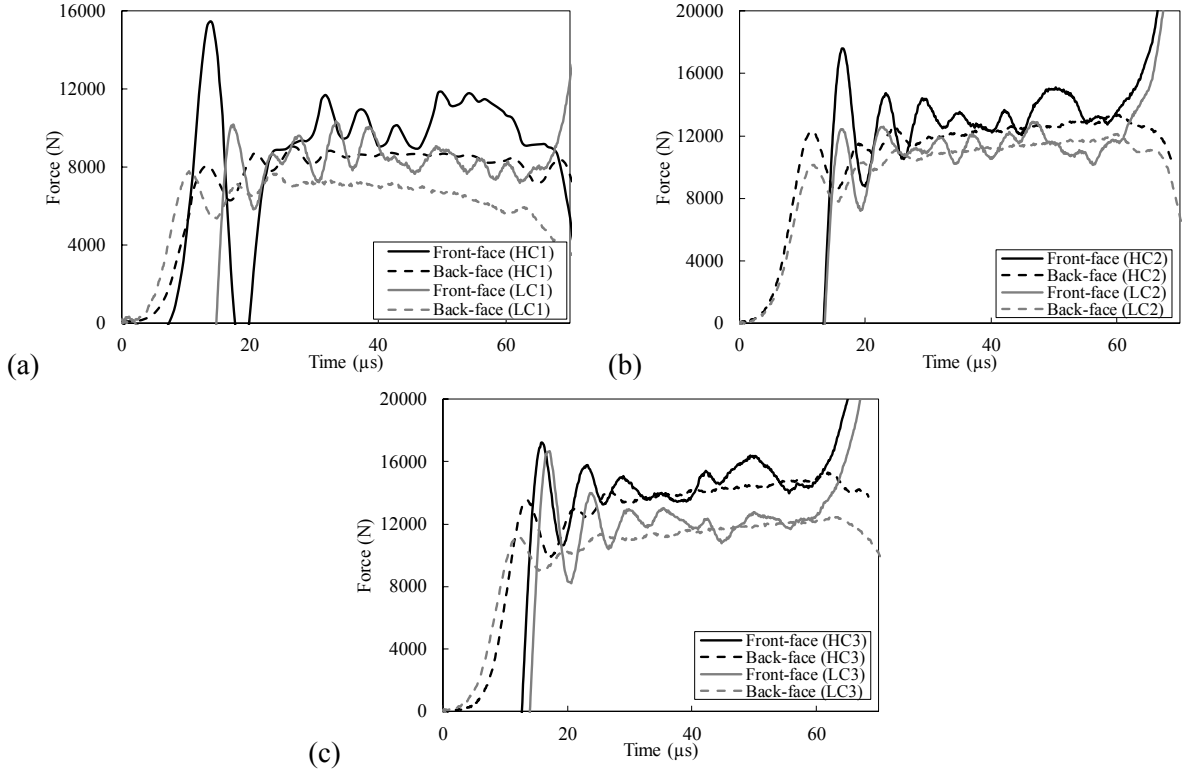

Figure 8: Measured force at the front and back faces for (a) HC1 and LC1, (b) HC2 and LC2, and (c) HC3 and LC3 specimens during the SHB test.

*4.4 Finite Element Simulation*

  Figure 9(a, b) show the results of FE simulations of the HC1 specimen under dynamic load where the coefficient of friction between the specimen and the top and bottom plates are assumed to be (i) equal and (ii) different, respectively. The friction coefficient between the two surfaces mainly depends on the surface finish quality (i.e. rough or finished) and the surface lubrication quality (dry or well-lubricated). The common friction coefficient for the first case (Figure 9(a)) is assumed to be 0.5, while, for the second case (Figure 9(b)) the coefficient of friction on the top and bottom surfaces are assumed to have different values of 0.01 and 1, respectively. It can be seen that the shape of the specimen for the first case is symmetric with respect to a plane parallel to the specimen ends crossing the middle of the specimen; while, the shape of the second case with different top and bottom friction coefficients looks similar to the shape of the specimen after SHB test. Different coefficients of friction on the top and bottom of the specimen is caused by the top surface of the specimen being smooth, while the bottom surface is rough, which is an inherent limitation and consequence of the AM process. As shown in Figure 4(c), the specimens bottom surfaces are attached to a support structure which is sawed after the AM process, that leaves the bottom surface rough. Although, to get a good surface finish at the bottom surface the specimen could be machined, to avoid damaging the struts during surface machining was not performed. A similar simulation is performed for all the specimens and similar behavior is observed (results are not presented for the sake of briefness). Since local buckling at one end of the specimens is also observed in the quasi-static tests, it can be concluded that it is not likely that the damage initiation at one end of the specimens in SHB experiments is merely due to the force imbalance.

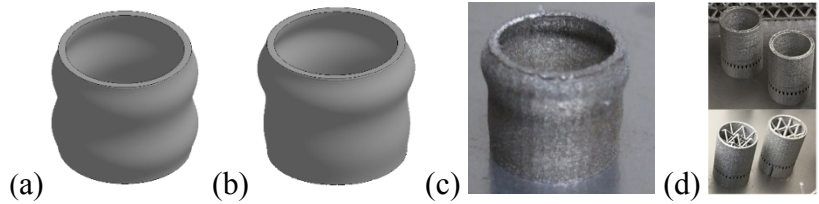

(a) (b) (c) (d)

Figure 9: FE results for deformation of HC1 specimen under dynamic loading with (a) equal and (b) different coefficient of friction on the specimen top and bottom surfaces; (c) photograph of the HC1 specimen after SHB test, and (d) photograph of the specimens on the support structure after AM build.

Figure 10(a, b) show the contours of plastic deformation in HC1 and LC1 specimens during dynamic loading, respectively. It can be seen that (i) up to t=10 $\mu s$ the specimen deformation is elastic and plastic strain is negligible, (ii) from 10 $\mu s$ to 40 $\mu s$ the deformation is (mainly) in the form of axial plastic deformation of the cylinder and struts, and (iii) after t=40 $\mu s$ the cylinder buckling occurs. Similar contours are plotted for specimen sets # 2 and 3 and it is observed that the deformation in these specimens remains mainly in the plastic regime (no buckling). Figure 11 shows the dynamic force-displacement curves obtained from the FEA for various specimens together with their final deformed shape. The elastic, plastic (hardening), and post-buckling (softening) regimes can be distinguished for HC1 and LC1 specimens in this figure, similar to what was observed in the experiments. It is also observed that the stiffness of the LC specimens is lower than their equivalent HC specimens for all the cases. In addition, the axial force is lower in the LC specimens compared to their equivalent HC specimens which is in agreement with experimental data.

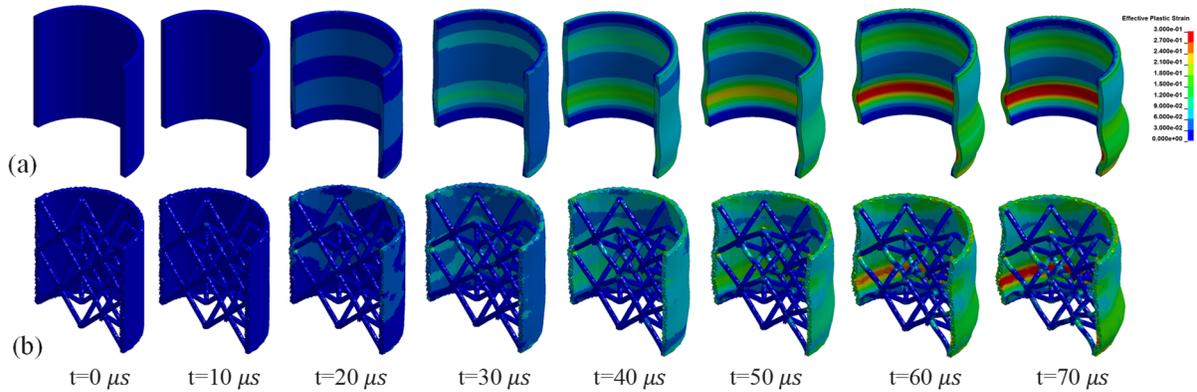

Figure 10: Contours of plastic strain for (a) HC1 and (b) LC1 specimens during dynamic loading.

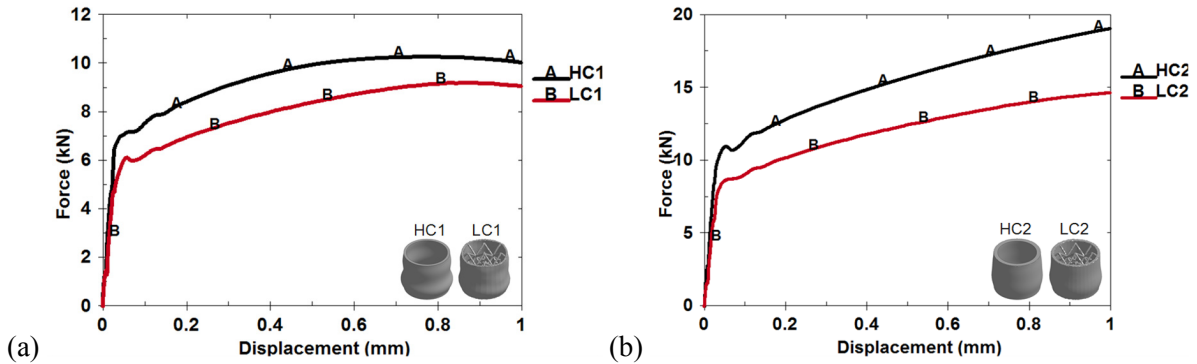

(a) (b)

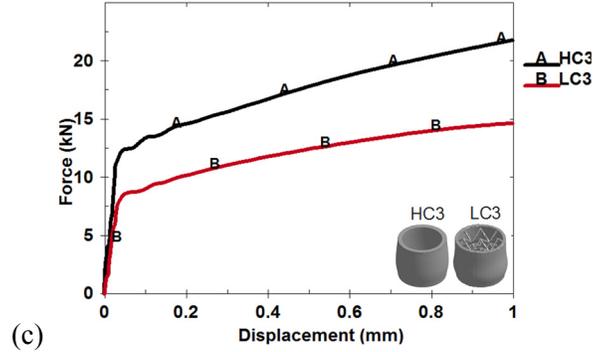

(c)

Figure 11: FEA force-displacement curves for (a) HC1 and LC1, (b) HC2 and LC2, and (c) HC3 and LC3 specimens.

Comparing the experimental force-displacement curves of Figure 7 with the FEA results of Figure 11, it can be observed that the force value in the FE simulations is higher than the corresponding values from the SHB experiments at the same displacement value. There are at least three possibilities for this difference (i) the boundary conditions in the FEA are different than the SHB test, where a percentage of the incident wave gets reflected into the incident bar, while the FE simulations do not take this into account, (ii) thermal softening is not considered in the simulations (material constant $m$ in the JC model is assumed to be zero), (iii) there are imperfections in the specimen which reduce material strength. As mentioned previously, modeling the bars explicitly in the SHB experiment is computationally expensive and is not performed here. To study the contribution of the thermal softening on the specimen behavior, the force-displacement curve of the HC1 specimen is plotted assuming the value of $m = 1$ in the JC model. Figure 12 shows the force-displacement results for two cases with $m = 0$ and $m = 1$. It can be seen that by considering the thermal softening effect in the simulations, the change in the force-displacement curve is negligible.

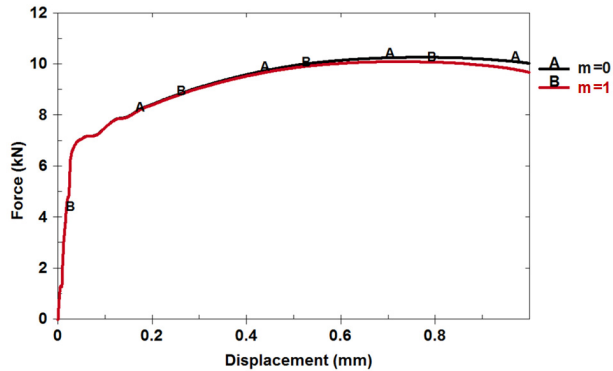

Figure 12: Effect of thermal softening coefficient $m$ on force-displacement curve of HC1 specimen.

## 5  Discussion

Figure 13(a) shows the comparison between the quasi-static and dynamic force-displacement curves for LC1 and HC1 specimen. It can be seen that the force value where the post-buckling softening occurs in the SHB test is higher than the corresponding quasi-static values. In addition, the displacement at which the softening initiates is lower in the SHB tests compared to the corresponding quasi-static values. Specifically, by increasing the loading rate from 2.5E-2 mm/s in the quasi-static test to 1.5E3 mm/s in the dynamic test (five orders of magnitude increase in the loading rate) about 36% and 29% increase in the force value where softening initiates is observed for LC1 and HC1 specimens, respectively. Since this behavior is observed in both HC1 and LC1 specimens it suggests that the increase in the measured force in dynamic tests compared to quasi-static tests is possibly due to the strain rate hardening of the bulk AM-SS316L material. To study this effect, FE simulation of the HC1 specimen is performed by setting the material parameter $C=0$ in order to remove the strain rate hardening effect from the simulation. The results for this simulation are given in Figure 13(b) which shows that the peak force for the case with $C=0.011$ is 6.5% higher than the case with $C=0$; while the displacement at which the peak force occur shifts to lower values. This behavior is qualitatively in agreement with what is observed experimentally in Figure 13(a).

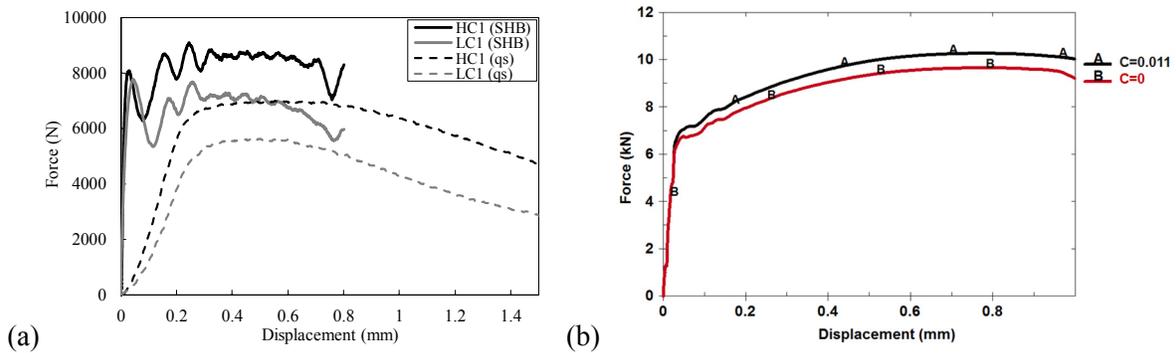

Figure 13: (a) Quasi-static and dynamic force-displacement curves for LC1 and HC1 specimens and (b) FEA force-displacement curve of HC1 for two different values of strain rate hardening coefficient $C$.

Table 5 shows the stiffness-ratio and peak force-ratio, defined by the ratio of the corresponding value for LC specimen over its equivalent HC specimen. It can be seen that for all the specimens the stiffness-ratio and peak force-ratio are less than one, which means that LC specimens have a lower stiffness and peak force compared to their equivalent HC specimens. The ratio of the stiffness-ratio to peak force-ratio for all the specimens is also given in this table, which provides a measure for the stiffness reduction vs. peak force reduction in the LC specimens compared to their equivalent HC specimens. It can be seen that for both quasi-static and dynamic regimes this ratio is less than one for all the specimens. This means that there is greater reduction in the stiffness compared to the reduction in the peak force for all the LC specimens when compared with their equivalent HC specimens. In the other words, the reduction in the peak force in the LC specimens comes at the expense of a greater reduction in the stiffness. Therefore, if a structure is intended to be used as a load carrying member with tailored impact energy absorption, the lattice architecture needs to be designed for optimal structural stiffness as well as energy absorption.

Table 5: Experimental stiffness-ratio and peak force-ratio for different specimens.

|  | Stiffness-ratio | Peak force-ratio | | Stiffness-ratio/peak force-ratio | |
| --- | --- | --- | --- | --- | --- |
|  | QS | QS | SHB | QS | SHB |
| LC1/HC1 | 0.70 | 0.81 | 0.87 | 0.86 | 0.80 |
| LC2/HC2 | 0.57 | 0.83 | 0.88 | 0.68 | 0.65 |
| LC3/HC3 | 0.68 | 0.70 | 0.81 | 0.97 | 0.84 |

Figure 14(a-c) show the strut and cylinder buckling in LC1 specimen obtained from quasi-static test, SHB test, and FE simulation, respectively. These figures show that struts fail in a similar mode in both quasi-static and dynamic tests. In addition, it can be seen that FE simulations could qualitatively model the mechanics of deformation of the specimen, especially, the pattern in which the struts and the cylinder buckle. It can be understood that the main mechanisms of energy absorption in the LC1 specimen are (i) axial plastic deformation of the cylinder and struts, (ii) buckling of the struts and cylinder, and (iii) transformation of the axial incident energy to radial deformation of the struts and the cylinder. Although, the cylinder axial plastic deformation and buckling also exist in the HC1 specimen, the additional energy absorption mechanisms provided by lattice structure are the main reasons for the additional reduction in the transmitted force observed in the LC1 specimen in the dynamic experiment. By developing simplified tools that are capable of modeling the mechanics of deformation of lattice structures and using optimization techniques one can tune these energy absorption mechanisms to tailor lattice architecture to maximize their specific energy absorption (per unit weight) for a given application.

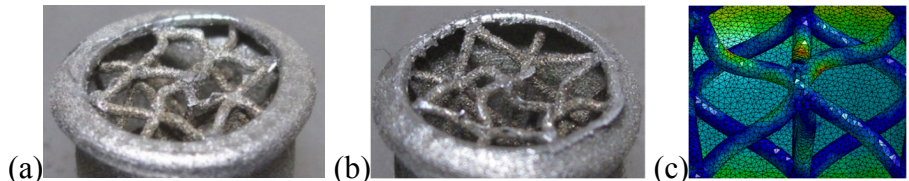

(a) (b) (c)

Figure 14: Strut and cylinder buckling of LC1 specimen observed in (a) quasi-static test, (b) SHB test, and (c) dynamic FE simulation.

## 6 Conclusion

Quasi-static and dynamic tests are performed on additively manufactured stainless steel lattice cylinders and their equivalent hollow cylinders with the same length, outer diameter, and mass. Lattice cylinders show a lower transmitted force compared to their equivalent hollow cylinders in SHB experiments. However, lower transmission in the lattice cylinders comes at the expense of a greater reduction in their stiffness. Lattice cylinders have additional energy absorption mechanisms which reduce the axially transmitted force by transforming it into lateral deformation. FE simulations are performed and it is observed that the simulations could qualitatively capture the overall behavior of the specimens. In particular, FE simulations are used to gain a better understanding of dynamic behavior of the lattice structure and to study the effect of strain rate hardening, thermal softening, and the friction coefficient between the specimen and the test apparatus. Three distinct regimes of (i) elastic, (ii) plastic (hardening), and (iii) post-buckling (softening) are observed in the force-displacement curves in quasi-static tests, dynamic tests, and FE simulations. In addition, it is observed that increasing the loading rate by five orders of magnitude leads to up to 36% increase in the peak force that the lattice cylinder can carry, which is attributed to strain rate hardening effect in the bulk stainless steel material.